# A Discovery Plan for Pharmacy Benefit Managers' Collusion

by
Lawrence. W. Abrams, Ph.D
October 31, 2024

## Summary

The Federal Trade Commission (FTC) has recently filed an administrative complaint against the Big 3 pharmacy benefit managers (PBMs) claiming they engaged in unfair conduct in violation of Section 5 of the FTC Act, 15 U.S.C. § 45. They never used the word "collusion" in the complaint and chose not to sue under The Sherman Act, Section 1. We view this as a novel case of market design collusion rather than a case of price collusion. The Big 3 PBMs are conceptualized as auctioneers soliciting rebate bids off unit list prices in exchange for favored positions on formularies.

We will show how the fairness standard of the FTC Act can be made operational by judging fairness against economic theories of good auction design. Discovery is focused on finding explicit communication among the Big 3 PBMs in 2012 to change the so-called "winner's determination equation" of this auction, adding high gross rebates as a basis for formulary position assignments. On the other hand, we will argue that a case based on a bevy of anecdotes comparing only net unit prices will fail due to complexities in the winner's determination equation.

## Disclosures:

I have not received any remuneration for this paper nor have I financial interest in any company cited in this working paper. I have a Ph.D. in Economics from Washington University in St. Louis and a B.A. in Economics from Amherst College. Other papers on PBMs can be found on my website https://nu-retail.com

## I. Introduction

The FTC has recently filed an administrative complaint against the Big 3 pharmacy benefit managers (PBMs) claiming they engage in unfair conduct in violation of Section 5 of the FTC Act, 15 U.S.C. § 45.[1] They never used the word "collusion" in this complaint and chose not to sue under The Sherman Act, Section 1.

This is not a typical case of price-fixing by a small group of sellers. While the FTC has plenty of evidence of shadow list pricing among pharmaceutical manufacturers (Pharma), there is also plenty of evidence indicating that this is a case of "conscious parallelism" by Pharma reacting independently to PBM suggestions for how best to structure bid strategies.

The FTC claims that the Big 3 PBMs as agents make formulary assignments not in the best interests of plan sponsors as principals, favoring high gross rebate drugs to the exclusion of the lowest unit net prices of therapeutic equivalents. In defense, the PBMs' economics consultants (hereafter the Carlton Report) argue that these comparisons are anecdotal.[2]

---

[1] Federal Trade Commission, *Complaint - Public Redacted Version*, Docket 9437, (September 20, 2024), https://www.ftc.gov/system/files/ftc_gov/pdf/d9437_caremark_rx_zinc_health_services_et_al_part_3_complaint_public_redacted.pdf

[2] Dennis W. Carlton, et. al., *PBMs and Prescription Drug Distribution: An Economic Consideration of Criticisms Levied Against Pharmacy Benefit Managers" Compass Lexecon,* (October, 2024), p. 114, https://compass-lexecon.files.svdcdn.com/production/files/documents/PBMs-and-Prescription-Drug-Distribution-An-Economic-Consideration-of-Criticisms-Levied-Against-Pharmacy-Benefit-Managers.pdf?dm=1728503869.

We view this as a novel case of market design collusion rather than a case of exchange price collusion. The Big 3 PBMs are conceptualized as auctioneers soliciting unit rebate bids off list prices in exchange for favored positions on formularies. At the highest level, we frame the case against PBMs as explicit collusion to change a key rule of this auction, starting a trend of copayment inflation and harm to patients covered by drug benefit plans.

We will show how the fairness standard of the FTC Act can be made operational by judging fairness against economic theories of good auction design. The "smoking gun" in this case are explicit communications among the Big 3 PBMs to change the so-called "winner's determination equation" of this auction by adding high gross rebates as a basis for formulary position assignments.

We want to be clear. The claim is that PBMs **added** gross rebates as a basis, not **substituted** gross rebates for net price after rebates. Most of the time, net prices outweighed gross rebates in assigning positions. Occasionally, the reverse was true. That is one among several other reasons detailed below why the FTC case based on anecdotal evidence is weak.

We pinpoint 2012 as the likely year the collusive communications began. The motivation was a need to replace retained rebates from blockbuster small molecule drugs losing patent protection with retained rebates from a smaller pool of self-injectable biologics.

Based on discovery, we see a hierarchy of cases against the Big 3 PBMs:

| Level | Case | "Smoking gun" | Conscious Parallelism "Plus Factors" | Plan Sponsorship |
|---|---|---|---|---|
| | **A Hierarchy of Cases Against the Big 3 PBMs to Change the Winners' Determination Equation of their Auction Design for Favored Formulary Positions** | | | |
| I | Explicit collusion with no plan sponsorship | Yes | Yes | No |
| II | Conscious parallelism with no plan sponsorship | No | Yes | No |
| III | Conscious parallelism with plan sponsorship | No | Yes | Yes |

## II. An Auction Market Design

PBMs started out as computer networking specialists who automated prescription claims processing by connecting pharmacy point of sales terminals to back-office health insurance mainframes. Sometime in the 1990s, PBMs added a look-up table to the point of sale software to automate switching of off-patent brands to lower cost generics.

Starting in 2000, the most popular therapeutic classes of drugs -- proton pump inhibitors, COX-2 inhibitors, 2nd generation antihistamines, and statins -- started to see the entry of therapeutic equivalents with a mere entamer difference in molecular structure. The opportunity for capturing some of the excess profits generated by patent-protected drugs was an order of magnitude greater than surplus captured from switching off-patent brands to generics.

Give PBMs credit for realizing that formularies could be used to create competition among drugs that otherwise were patent protected monopolists. At the same time, it became problematic that they were health plan agents with no discretion over formulary management and therefore not subject to fiduciary laws. It also became problematic that they were aligned with goals of plan sponsors as the PBM opaque reseller business model became dependent on retained rebates.

Our contribution to the PBM debate started in 2004, when we first disaggregated PBM 10-Q financials showing their dependence on retained rebates.[3] Lately, we have begun to apply the economics field of market design to better understand the exchange of rebates for favored formulary positions.[4]

We had an "aha" moment in 2023 when we read sections of The Grassley-Wyden Senate Staff Report (hereafter the Grassley-Wyden Report) based on over 100,000 pages of internal documents and emails relating to rebate negotiations between the Big 3 PBMs -- CVS Caremark, Aetna Express Scripts, and United Healthcare OptumRx -- and the Big 3 Insulin drug manufacturers -- Sanofi, Novo Nordisk, and Eli Lilly.[5]

---

[3] Lawrence W. Abrams, *Papers on the PBM Business Model*, (2003 - 2019), https://www.nu-retail.com/pbm-business-model.

[4] Lawrence. W. Abrams, *The Market Design for Formulary Position,* (March 2023), https://c1c0481a-8d34-49dc-ad66-d6c488c905a9.usrfiles.com/ugd/c1c048_a0a9df7cf0de4a1094872ebd0f390941.pdf.

[5] United States Senate Finance Committee Staff Report, Charles E. Grassley, Chairman & Ron Wyden, Ranking Member, *Insulin: Examining the Factors Driving the Rising Cost of a Century Old Drug Staff Report,* (January 2021), https://www.finance.senate.gov/imo/media/doc/Grassley-Wyden%20Insulin%20Report%20(FINAL%201).pdf

Three key features of these negotiations caught our eye. First, PBMs offered a "menu" of bid options featuring both exclusive position and shared positions. Second, the interactions were multi-round offers followed by suggestions to raise bids in order to avoid exclusion. Third, the basis of rebate offers was expressed as a % off unit wholesale acquisition costs (WAC).

Based on a taxonomy of auction designs, we have conceptualized this exchange as a multi-round combinatorial auction.[6] It is doubtful that PBMs have ever labeled their annual negotiations with Pharma as an auction. Nevertheless, it is an exchange, or market, with a particular design featuring mutually agreed upon rules. It has been repeated yearly now for over 20 years. We think it is reasonable to view it as a "vernacular" auction design that works, just as scholars have found that a lot of vernacular buildings designed centuries ago are consistent with modern architectural design theory.

### III. The Winners' Determination Equation

The purpose of this section is to show that the introduction of shared formulary positions to the menu of bid options complicates the winner's determination equation. So much so, we believe that a discovery plan for PBM collusion should focus on explicit communication to collude on auction design. This is in contrast to the FTC's discovery plan based only on comparisons of net unit prices after rebates.

---

[6] Nikhil Agarwal & Eric *Budish, Market Design*, National Bureau of Economic Research Working Paper Series, No. 29367 (October 2021), p. 7-8, http://www.nber.org/papers/w29367.pdf.; Peter Crampton, et. al, COMBINATORIAL AUCTIONS, MIT Press, (2006), Introduction. https://cramton.umd.edu/ca-book/cramton-shoham-steinberg-combinatorial-auctions.pdf

We start by proposing three features of this auction in addition to those that jumped out at us from the Grassley-Wyden report. This includes (1) a common value auction; (2) a bid menu featuring shared formulary positions with varying numbers of assignments; (3) the winner's determination equation.

Based on a taxonomy of auction designs, the PBM auction would be classified as a common value auction.[7] This is because Pharma's willingness to pay for formulary positions is profitability, a common value that both sides of this exchange can estimate. PBMs have chosen a bid basis expressed as % off publicly available drug list unit prices known as the wholesale acquisition costs (WAC). This is good auction design in terms of encouraging bidding. Unit bids versus full lump sum bids reduces risks of overestimation of profitability and the "winner curse" associated with common value auctions.

Fortuitously, choosing the publicly available WAC as part of the bid basis has made it easier for PBMs to collude and enforce a change to the two step high list - high rebate design. Without publicly available WAC as part of the basis, shadow list pricing by Pharma would have been difficult to monitor by both Pharma and PBMs.

Allegations of PBMs' unfairness consider only a winner's determination equation for a single position in any formulary therapeutic class for a single year. Normally, the objective of an auctioneer is to maximize proceeds for its principal, or in this case minimize total drug benefit costs. If only exclusive positions were up for auction, then a fair assignment would be to award exclusive positions to that Pharma with the lowest net unit price after unit rebates, regardless of gross rebates or expected market share.

---

[7] Crampton, et. *al., supra* p.33.

There are both anticompetitive and procompetitive reasons why PBMs make assignments using bases in addition to net unit prices after rebates. The anticompetitive reason stems from the motive of the alleged collusion -- adding gross rebates as a basis was designed to fuel list price inflation and reestablish the upward trend in retained rebates dollars.

We want to be clear. The allegation is that PBMs **added** gross rebates as a basis, not **substituted** gross rebates for net price after rebates. Most of the time, net prices outweighed gross rebates in the winners' determination equation. Occasionally, the reverse was true. Finding what tipped the scales one way or another would require a statistical study involving a large sample size of assignments accompanied by all competitors' bid menus. In addition, a pharmacological profile of each drug and the year it came to market should be factored in.

Even if gross rebates were not a basis, there are both anticompetitive and procompetitive reasons why PBMs make assignments other than just on the basis of net unit prices. The explanation starts with good auction design which dictates a bid menu of both exclusive and shared positions. The question is what is the auction design rationale for shared positions and what other bases do PBMs use in the winners' determination equation.

First, offering a shared position option is equivalent to "set aside" bid packages in government procurement auctions designed to nurture emerging growth companies and long term competition. Like government procurement auctions, the PBM auction is repeated each year. The goal of PBMs and their plan sponsors is not just single year drug benefit cost minimization, but long term minimization.

It is in the best interest of PBMs and their plan sponsors to nurture competition by offering shared position options to new entrants to a therapeutic class.. However, we have shown in another paper that shared positions in "immature therapeutic classes" where market share of entrants is uncertain complicates the winner's determination equation.[8]

The second rationale is that PBMs recognize that therapeutic equivalents are not perfect substitutes. Offering shared positions in any given therapeutic class can be viewed as a recognition by PBMs of the need to satisfy physician and patient preferences for some choice over strict adherence to total benefit cost minimization.

Based on auction design theory, there is a third reason. Offering a combinatorial menu of bid packages improves bid elucation as they capture the superadditive or subadditive value of combinations.[9] The textbook example is an auction that offers a bid menu of airplane tickets to Hawaii, a week's stay in Hawaii, and a third package that combines both in order to capture the superadditive value of airplane tickets + hotel room for some bidders.

In the PBM case, the profitability of a shared formulary position is subadditive as the number of allowed assignments increases. The reason is that average unit profitability of a shared position decreases with the number of assignments due to increasing average unit production and marketing costs. Generally, Pharma's willingness to pay as expressed in % off WAC bids is lower for shared positions than for an exclusive position. This negative relation between unit rebate bids and the number specified in a shared position bid package has been substantiated by the Grassley-Wyden Report.[10]

---

[8] Abrams,*The Market Design for Formulary Position, supra*.
[9] Crampton, *supra* p.36.
[10] Grassley-Wyden, *supra* p.40.

There is a fourth complication that is a consequence of only requiring a unit bid basis for rebates. To recap, this is a common value auction where willingness to pay is based on estimated profitability of any given position assignment. Profitability is the product of unit margin dollars times expected quantity demand of any given assignment. Requiring only a bid basis in units rather than lump sums is good auction design because it reduces bidder risk and encourages bid elucidation. But, assigning positions only on the basis of net unit bids creates the possibility that favored positions go to drugs with relatively little demand.

Our insight into this complication came from an awareness of Google's ad position auction experience. It was Google's chief economist Hal Varian who introduced Google to the economics of position auctions.[11] The winners' assignment equation required both $ per click and an estimate of click-through rates. Assigning ad positions only on $ per click could result in too many top positions going to high unit bidders with ads having little appeal and low click-through rates.

It was Google's co-founder Larry Page who developed a complex estimate of expected click-through rates called AdRank. Here is a link to a video of Google's chief economist Hal Varian presenting an example of how Google's position auction works.[12] In the example, the top ad position went to the bidder with a relatively low unit bid but a very high AdRank.

---

[11] Hal Varian, *Position Auctions*, " INTERNATIONAL JOURNAL OF INDUSTRIAL ORGANIZATION 25 (2007) 1163-1178, https://people.ischool.berkeley.edu/~hal/Papers/2006/position.pdf.

[12] Hal Varian, *Pay Per Click Management - Insights on the Google AdWords Auction System,* Youtube video, (October 24, 2014), https://www.youtube.com/watch?v=tW3BRMld1c8

The need by Google to go beyond simple $ per unit bids is similar to the need for PBMs to go beyond assigning formulary positions solely on the basis of net unit prices. Because the objective of a PBM is to minimize total benefit costs, this requires PBMs to assign formulary positions on the combined bases of net unit prices and estimates of expected demand. In another paper, we present a linear algebraic model of the winner's determination equation in this case.[13]

In sum, there are lots of nuances in the PBM winner's determination equation. A discovery plan based only on comparison of net unit prices is problematic.

### IV. Motive

The Carlton Report fails to consider changes in PBMs' distribution of gross profits by source over time and how that might be a factor in this case. As evidence of no PBM motive, the Carlton Report presents a graph showing that PBM gross profit margins have remained in the 8% range over time.[14] However, they conveniently left out a graph showing an unabated rise in gross profit dollars even though margins have remained constant.

Both the FTC administrative complaint and the Carlton Report rebuttal are weak on timelines. The Carlton Report only offers generalities about PBMs offering contracts specifying options. On the other hand, The FTC inexplicitly weakens its case by offering ample evidence since 2019 of PBMs offering plan sponsors contracts specifying options for formularies with different orientations around list prices, gross rebates, copayments and rebate guarantees.[15]

---

[13] Abrams, *The Market Design for Formulary Position, supra.*
[14] Carlton*, supra*, p. 75.
[15] Federal Trade Commission, *supra,* paragraphs 104-8.

The lack of discussion on both sides of the timeline of PBM conduct is all the more glaring given clear evidence first compiled by IQVIA, and made public by Adam Fein, of a gross-to-net bubble starting in 2013.[15] Bidding for favored positions starts about 3-5 months before being made public in December with a January 1st effective date. As a result, we place 2012 as the likely year PBMs first communicated to Pharma a change in the winner's determination equation. We also pinpoint 2012 as the year to discover explicit collusion among the Big 3 PBMs to make this change.

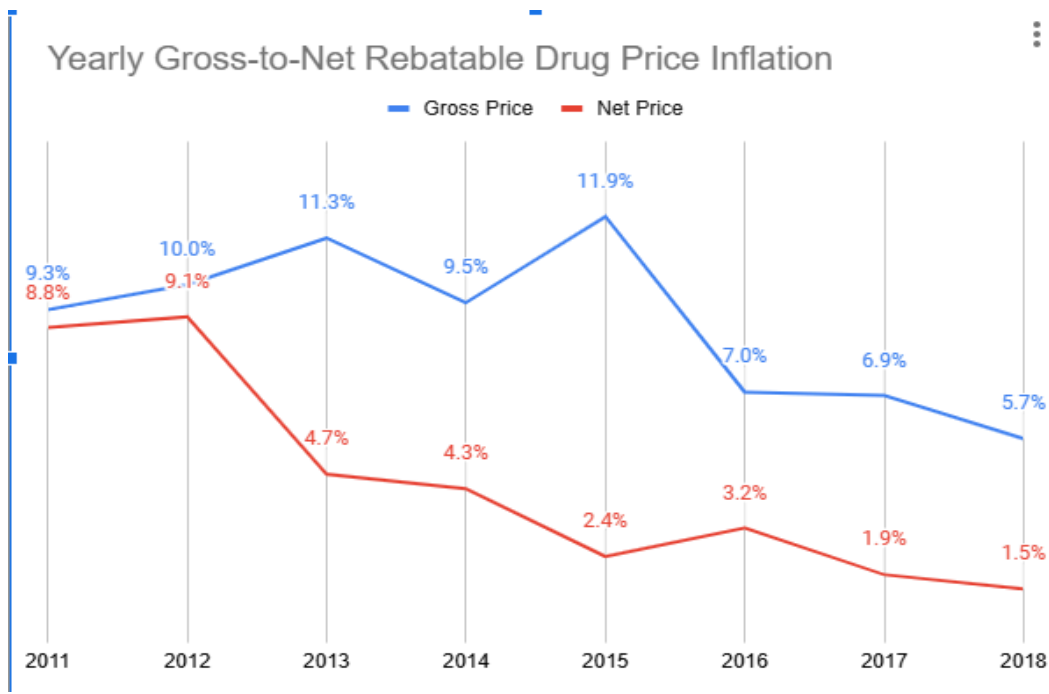

Data from A.Fein blog posts of IQVIA data,, 2017 and 2019, supra.

---

[15] Adam Fein, *Drug Channels*, blog post, (June 14, 2017). https://www.drugchannels.net/2017/06/new-data-show-gross-to-net-rebate.html.
 Adam Fein, *Drug Channels*, blog post, (January 29, 2019). https://www.drugchannels.net/2019/01/drug-prices-are-not-skyrocketingtheyre.html

On the surface, it looks like Pharma initiated the bubble. In 2017, we presented the case that it was initiated by PBMs motivated by a need to replace declining retained rebates from blockbuster small molecule drugs losing patent protection with retained rebates from a much smaller group of rebatable self-injectable biologics.[16]

Plus, the early 2010s was the beginning of push back by plan sponsors asking for contract specifications to limit rebates to a certain percent or dollar amount.  So, PBMs independently concluded that the only way they could increase retained rebate dollars over time without increasing the rebate retention rate would be to motivate Pharma to embark on yearly list price inflation, especially self-injectable biologics like insulin.

## V. The Role of Plan Sponsors

As we noted in our introduction, we see a hierarchy of allegations against PBMs. At the highest level, the allegation is that PBM explicitly colluded without plan sponsor awareness to change in the winners' determination equation and related high copayments. Next is the allegation that the change was not explicit collusion but conscious parallelism. Again, this is without awareness or sanction by plan sponsors.

The third is the Carlton Report allegation that PBMs offer plan sponsors contract options around formulary designs, copayments, rebate retention and PBM compensation. Thus, it is plan sponsors as principals, not PBMs as agents, who should be held accountable for such changes.

---

[16] Lawrence. W. Abrams, *Blame Pharmacy Benefit Managers Not Pharma for Driving Drug Price Inflation,* (September, 2017). https://c1c0481a-8d34-49dc-ad66-d6c488c905a9.usrfiles.com/ugd/c1c048_661c6fd136b44a3880a0f2d0f3f8baa9.pdf

To avoid sinking to the lowest level of claims against PBMs -- collective parallelism with "plus factors" with plan sponsorship -- a discovery plan should include a counterfactual search of communications between PBMs and plan sponsors in 2012 regarding adding gross rebates as a basis for formulary assignment.[18]

We end with this final note. There will be unintended consequences of this high profile case. This is because the case is bringing to the forefront the question of due diligence by plan sponsors as fiduciaries under The Employee Retirement Income Security Act (ERISA). The recent class action suit initiated by an employee of Johnson & Johnson claiming breach of fiduciary duty as the plan sponsor of employees' drug benefit plan could be the beginning of a tidal wave of civil lawsuits against companies.[19]

---

[18] William E. Kovacic, *Antitrust Policy and Horizontal Collusion in the 21st Century* Loyola Consumer Law Review Volume 9 | Issue 2 Article 13 1997.
https://lawecommons.luc.edu/cgi/viewcontent.cgi?referer=&httpsredir=1&article=1484&context=lclr

[19] Ann Lewandowski, on behalf of self and others vs Johnson and Johnson, et. al., *Class Action Complaint*, Case 1:24 - CV00671, United States District Court for the District of New Jersey filed 02/05/24.
https://fingfx.thomsonreuters.com/gfx/legaldocs/znpnkkrmbvl/EMPLOYMENT_JANDJ_ERISA_complaint.pdf